\documentclass[prl,aps,onecolumn,showpacs,preprintnumbers,amsmath,amssymb]{revtex4}

\usepackage{graphicx}
\usepackage{dcolumn}
\usepackage{bm}
\usepackage{color}

\begin{document}


\title{Accurate numerical verification of the instanton method for macroscopic quantum tunneling:
dynamics of phase slips}
\author{Ippei Danshita$^{1,2}$}
\author{Anatoli Polkovnikov$^{1}$}
\affiliation{
{$^1$Department of Physics, Boston University, Boston, MA 02215, USA}
\\
{$^2$Department of Physics, Faculty of Science, Tokyo University of Science, Shinjuku-ku, Tokyo 162-8601, Japan}
}

\date{\today}

\begin{abstract}
Instanton methods, in which imaginary-time evolution gives the tunneling rate, have been widely used for studying quantum tunneling in various contexts. Nevertheless, how accurate instanton methods are for the problems of macroscopic quantum tunneling (MQT) still remains unclear because of lack of their direct comparison with exact time evolution of the many-body Schr\"odinger equation. Here, we verify instanton methods applied to coherent MQT. Specifically applying the quasi-exact numerical method of time-evolving block decimation to the system of bosons in a ring lattice, we directly simulate the real-time quantum dynamics of supercurrents, where a coherent oscillation between two macroscopically distinct current states occurs due to MQT. The tunneling rate extracted from the coherent oscillation is compared with that given by the instanton method. We show that the error is within 10$\%$ when the effective Planck's constant is sufficiently small. We also discuss phase slip dynamics associated with the coherent oscillations.

\end{abstract}

\maketitle
Tunneling is one of the most fundamental concepts derived from quantum theory and is essential
 for understanding enormous variety of phenomena in different fields of physics, such as high energy, condensed matter, and atomic physics. The list of such phenomena includes the $\alpha$-decay of nuclei~\cite{razavy}, tunneling between vacuum states in quantum cosmology~\cite{coleman1,coleman2} and chromodynamics~\cite{thooft,vainstein, rajaraman}, macroscopic quantum tunneling (MQT) in quantum gases~\cite{ueda,anatoli2} and condensed matter~\cite{leggett-kagan,takagi}, and also includes potential applications in quantum information~\cite{chiorescu}.

Instanton methods are general schemes describing quantum tunneling within a semiclassical approximation~\cite{vainstein,rajaraman,sakita}. 
They are applicable to the broad range of problems listed above.
These methods are based on the solution of the classical equations of motion in imaginary-time 
coordinate allowing one to obtain an analytical expression for the tunneling rate. 
The instanton methods are closely related to the Langer's formalism of decay of metastable states 
due to thermal fluctuations~\cite{langer}. 
Given the versatility and utility of the instanton methods, it is important to examine how accurately 
they predict the actual tunneling rate. 
We note that this question is not entirely trivial. For example, for applicability of the Langer's formalism 
it is important that the thermal bath (which can be a part of the macroscopic system) is big enough 
to provide sufficient energy necessary to overcome the barrier separating metastable and stable 
phases. 

For single particle problems, especially in one-dimension,  the instanton methods can be easily verified because the corresponding Schr\"odinger equation can be solved numerically with arbitrary precision. On the other hand, such numerical verification of the instanton methods is usually very difficult for complex systems consisting of many degrees of freedom, such as MQT and tunneling decay of the false vacuum. 
Alternatively, in the context of the current-biased Josephson junction, where the phase difference between the two superconductors is regarded as a macroscopic quantum variable, experiments have been extensively compared with the theory of MQT~\cite{devoret,li}. 
It has been shown that the experiments and the theory are in agreement to the extent that the 
instanton method provides an estimate of the order of the magnitude of the tunneling rate.
However, this comparison is inevitably limited by the experimental uncertainty, which arises from 
the fact that the theory uses phenomenological parameters extracted from separate experiments.  
For rigorous verification of the instanton methods, it is necessary to make direct comparison of 
their predictions with the first principles many-body simulations in a complex system.

In this work we study MQT of supercurrents of bosons in a one-dimensional (1D) ring lattice to 
examine the validity of the instanton method applied to MQT. Using the time-evolving block 
decimation (TEBD) method~\cite{vidal1,danshita1}, we perform first principles simulations of the 
real-time dynamics of the corresponding Bose-Hubbard model. 
In the regime where the energies of two macroscopic states with different winding numbers are degenerate, we show that the supercurrent exhibits coherent oscillations. 
These oscillations are accompanied by phase slips which result in sudden change of the winding 
number characterizing the supercurrent. 
The tunneling rate is accurately extracted from the period of oscillations, while it is also 
calculated by the instanton method in the quantum rotor limit corresponding to large filling 
factors~\cite{anatoli2, sachdev}. 
We are thus able to compare the numerical TEBD results with the prediction of the instanton method 
with no ambiguity. 
Our main finding is that the error of the instanton method is within 10\% when the effective Planck's 
constant is sufficiently small. 
We also find that the coherent oscillations of current persist even between the degenerate states 
with winding numbers different by two. Such process corresponds to the dynamics associated 
with coherent oscillations of double phase-slips.

\section{Model}
We consider a system of $N$ bosons at zero temperature confined in a homogeneous
1D ring lattice of $L$ sites.
Recently, such a system has been experimentally realized in the context of quantum
gases~\cite{henderson}.
We assume a sufficiently deep lattice so that the tight-binding approximation is valid.
Then, the system is well described by the Bose-Hubbard model~\cite{fisher},
\begin{eqnarray}
\hat{H} = -J\sum_{j=1}^{L}( e^{-i\theta} \hat{b}^{\dagger}_j \hat{b}_{j+1} + {\rm h.c.})
            + \frac{U}{2}\sum_{j=1}^{L}\hat{n}_j(\hat{n}_j-1).
\label{eq:BHH}
\end{eqnarray}
where $\hat{b}_{L+1}\equiv \hat{b}_{1}$, reflecting the periodic nature of the ring lattice.
The field operator $\hat{b}^{\dagger}_j$ ($\hat{b}_j$) creates (annihilates) a boson on the
$j$-th site, and $\hat{n}_j$ is the number operator. $J$ is the hopping energy and $U$ the onsite interaction.
The phase twist $\theta$ can be controlled by rotating the lattice~\cite{hallwood,ana} (or equivalently by writing the Hamiltonian in the rotating frame). 
In the case of commensurate fillings, where the filling factor $\nu \equiv N/L$ is integer,
the Bose-Hubbard model exhibits a quantum phase transition from a superfluid to a Mott insulator 
as $U/J$ is increased.
Since our interest is in the dynamics of supercurrents, we focus only on the superfluid regime throughout this paper.

\section{Supercurrent dynamics}
For pursuing our main goal of examining the validity of the instanton method, it is imperative
to reveal basic properties of the quantum dynamics associated with Eq.~(\ref{eq:BHH}). In this section, we confirm that a supercurrent flowing through the ring lattice actually exhibits MQT during the real-time evolution.
In the next section, we will compare the period of oscillations extracted from the MQT dynamics with that obtained by the instanton method.

To treat the quantum dynamics,
we use the quasi-exact numerical method, TEBD~\cite{vidal1}, which is conceptually
equivalent to the well-known time-dependent density matrix renormalization group~\cite{white,schollwock}. 
This method allows us to compute accurately the evolution of many-body wave functions of 1D 
quantum lattice systems. 
Recently, TEBD has been successfully adopted  by one of us to a system with periodic boundary
conditions~\cite{danshita1}. 
In order to study how a supercurrent behaves as a function of time, one needs to prepare a current-carrying state as an initial state of the real time evolution. 
For this purpose, setting $\theta=\theta_0\equiv 2\pi n/L$, we first perform the imaginary time 
propagation for Eq.~(\ref{eq:BHH}), which provides a current-carrying state with the winding 
number $n$. 
At $t=0$ the phase twist is suddenly shifted to $\theta_1\equiv \pi(2n-m)/L$, $m\geq 1$, so that another 
state with the winding number $n-m$ is exactly degenerate with the initial state. 
We then simulate the dynamics in this system propagating the initial state in real time. 
First we analyze the situation where the filling factor and the initial winding number are equal to 
unity: $\nu=1, n=1$ and investigate how the time evolution of supercurrents  depends on the 
parameters of the model $U/J$ and $\theta_1$.

\begin{figure}[tb]
\includegraphics[scale=0.45]{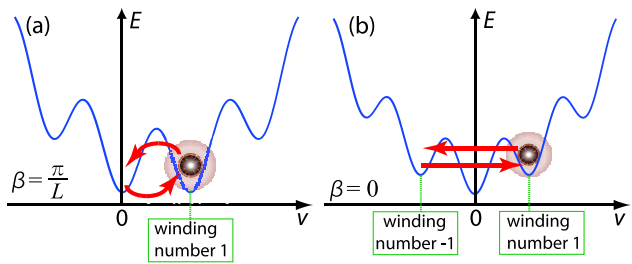}
\caption{\label{fig:MQT}
Sketch of the quantum dynamics of supercurrents in the effective potential obtained from the Bose-Hubbard Hamiltonian
Eq.~(\ref{eq:BHH}) for $\theta_1 = \pi/L$ (a) and $\theta_1 = 0$ (b). The blue solid lines sketch the energy landscape versus the current velocity $v$. The black circles represent the quantum state. Note that the plot represents a sketch of an actual process occurring in the multi-dimensional phase space.
}
\end{figure}
Let us start with the simplest case, $\theta_1= \pi / L$, where two macroscopically distinct states with winding numbers 1 and 0 are the degenerate lowest energy states. This situation is analogous to a superconducting flux qubit realized in a superconducting quantum interference device (SQUID), where two flux states with different winding numbers
are degenerate producing coherent Rabi oscillations~\cite{chiorescu}. Likewise in our case we expect coherent oscillations between the two degenerate states via MQT as sketched in Fig.~\ref{fig:MQT}(a).

\begin{figure}[tb]
\includegraphics[scale=0.42]{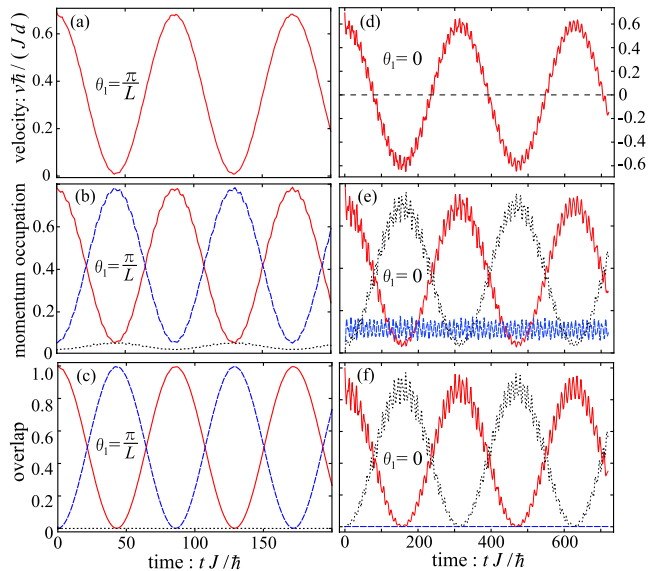}
\caption{\label{fig:CON}
Time evolution of the current velocity $v(t)$ (a) and (d), the momentum occupation $n(p,t)$
(b) and (e), and the overlap $|\langle \Phi_p |  \Psi(t) \rangle|^2$ (c) and (f).
For  $n(p,t)$  and $|\langle \Phi_p |  \Psi(t) \rangle|^2$, the red solid, blue dashed, and black
dotted lines correspond to $p=1, 0$, and $-1$.
We set $L=N=16$ and $U/J=2.5$. In (a), (b), and (c), $\theta_1 = \pi/L$, while in (d), (e), and (f) $\theta_1 = 0$.
}
\end{figure}

To demonstrate this, we first calculate the time evolution of the current velocity $v$ given by
\begin{eqnarray}
v=\frac{Jd}{i\hbar N} \sum_{j} \langle\hat{b}_j^{\dagger}\hat{b}_{j+1} - \rm{h.c.} \rangle,
\end{eqnarray}
where $d$ is the lattice spacing. When $U/J \ll 1$, the velocity is almost constant, i.e., the supercurrent is persistent. In contrast, when $U/J$ is sufficiently large, e.g. $U/J=2.5$, quantum fluctuations are strong enough to kick the state out from the one of the minima, and the superfluid coherently oscillates between the states with velocities $v(t=0)$ and $0$ as shown in Fig.~\ref{fig:CON}(a). The period of these oscillation decreases monotonically with $U/J$.

To confirm that these oscillations are due to quantum tunneling between two macroscopically distinct
states, we next calculate the overlap $|\langle\Phi_n|\Psi(t)\rangle|^2$ of the wave function with
the ground state $|\Phi_n\rangle$ of the Hamiltonian (\ref{eq:BHH}) with $\theta=2\pi n/L$, and
the momentum occupation $n(p,t)=\langle \hat{b}_{p}^{\dagger}\hat{b}_p\rangle$,
where $\hat{b}_{p}=L^{-1/2}\sum_{j}\hat{b}_j e^{-i2\pi p j/L}$.
In Fig.~\ref{fig:CON}(c), we show the overlaps with $|\Phi_1\rangle$, $|\Phi_0\rangle$, and
$|\Phi_{-1}\rangle$. The overlaps $|\langle\Phi_1|\Psi(t)\rangle|^2$ and $|\langle\Phi_{0}|\Psi(t)\rangle|^2$ are well approximated by the time dependence $\cos^2(\pi t/T)$ and $\sin^2(\pi t/T)$, respectively, where $T$ is the period of oscillations. Hence, the wave function is approximated by a macroscopic superposition of the states with
$n=1$ and $n=0$ (Schr\"odinger cat state) as
\begin{eqnarray}
|\Psi(t)\rangle \simeq \cos\left(\frac{\pi t}{T}\right) |\Phi_1\rangle
                                     + i \sin\left(\frac{\pi t}{T}\right) |\Phi_{0}\rangle.
\label{eq:cat}
\end{eqnarray}
In Fig.~\ref{fig:CON}(b), we show the momentum occupations for $p=1, 0, -1$, which behave almost identically to the overlaps, again justifying validity of the cat state description. We note that the similar cat state dynamics has been found also for quantum vortices in anisotropic traps~\cite{gentaro} and supercurrents in two-color optical lattices~\cite{andreas}.

We next consider the case of $\theta_1 = 0$, where $|\Phi_1\rangle$ and $|\Phi_{-1}\rangle$ are
degenerate. In this case, there are two possible scenarios of the fate of the supercurrent: (i) The supercurrent decays towards the zero momentum state creating excitations. (ii) It coherently oscillates between $|\Phi_1\rangle$ and $|\Phi_{-1}\rangle$ as sketched in Fig.~\ref{fig:MQT}(b). 
Previous theoretical work on the supercurrent decay anticipated the first scenario to calculate 
the lifetime of the metastable state using the instanton method~\cite{anatoli2,nishida,freire}. 
It is very likely that this scenario is indeed realized when the differences in winding numbers 
of $\theta_1$ and $\theta_0$ is large. 
In contrast, it is found in our numerical simulations that the second scenario mainly dictates 
the supercurrent dynamics as seen in Figs.~\ref{fig:CON}(d) and (f). 
The supercurrent exhibits a coherent oscillation between states with velocities $v(t=0)$ and $-v(t=0)$ 
with rapid wiggles. 
If these wiggles are ignored, then the wave function is well approximated by superposition of the 
states $|\Phi_1\rangle$ and $|\Phi_{-1}\rangle$. 
The zero momentum occupancy $n(p=0,t)$ (blue dashed line in Fig.~\ref{fig:CON}) oscillates in time 
with the same frequency as the wiggles in the $n(p=\pm 1,t)$ while the overlap of $|\Psi(t)\rangle$ 
with $|\Phi_0\rangle$ always remains zero. 
This means that the wiggles come from the coupling with the excited states with winding number 0
and that such states contribute to the wave function in addition to $|\Phi_1\rangle$ and 
$|\Phi_{-1}\rangle$.

\begin{figure}[tb]
\includegraphics[scale=0.47]{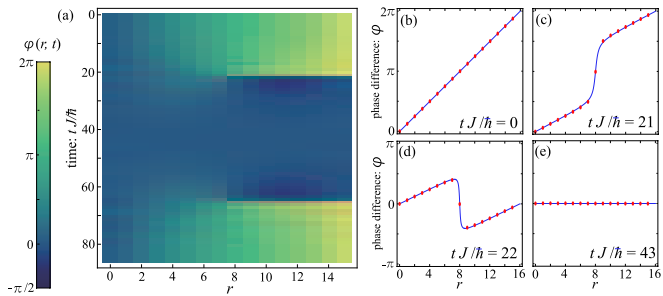}
\caption{\label{fig:single}
(a) Time evolution of the average phase difference $\varphi(r,t)$ for $L=N=16$,
$U/J=2.5$, and $\theta_1=\pi/L$. The phase jumps by $2\pi$ at the boarders between the bright and dark regions.
(b)-(e) Snap shots of $\varphi(r,t)$  for several values of $t$.
}
\end{figure}
\begin{figure}[tb]
\includegraphics[scale=0.47]{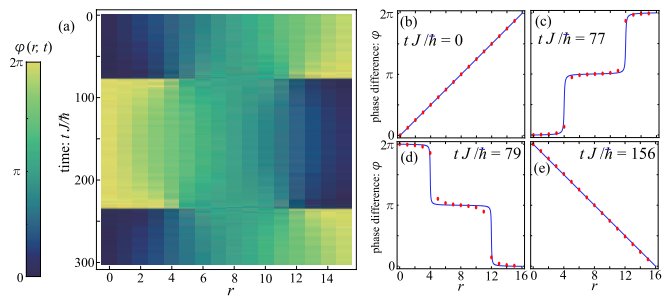}
\caption{\label{fig:double}
(a) Time evolution of the average phase difference $\varphi(r,t)$ for $L=N=16$,
$U/J=2.5$, and $\phi_1=0$. (b)-(e) Snap shots of $\varphi(r,t)$ for several values of $t$.
}
\end{figure}
Since during the coherent oscillations between the two degenerate states, the winding number
changes from 1 to 0 (or to $-1$), one expects emergence of the phase slip associated with these oscillations.
To reveal the phase slips, we calculate the time evolution of the average phase difference between the $j$-th and $(j+r)$-th sites, $\varphi(r,t)=\arg(\langle \hat{b}_j^{\dagger} \hat{b}_{j+r} \rangle)$. Notice that the phase difference is independent of $j$ because of the homogeneity of the system. In Fig.~\ref{fig:single}, we show $\varphi(r,t)$ that corresponds to the dynamics depicted in Figs.~\ref{fig:CON}(a)-(c). At $t=0$ (Fig.~\ref{fig:single}(b)), $\varphi(r,t)$ linearly changes with $r$  as $\varphi(r,t)=2\pi r/L$ corresponding to the winding number $n=1$. As time evolves, a phase kink develops around $r=L/2$ and it becomes $\sim \pi$ at $t=T/4$ (Fig.~\ref{fig:single}(c)). Immediately after $t=T/4$, the phase jumps by $2\pi$ and the winding number changes to $n=0$ as seen in Fig.~\ref{fig:single}(d).

In Fig.~\ref{fig:double}, we show $\varphi(r,t)$ that corresponds to the situation shown in Fig.~\ref{fig:CON}(d)-(f), where the supercurrent oscillates between the states with $n=1$ and $n=-1$. As $t$ increases, two phase kinks develop; which are localized around $r=L/4$ and $r=3L/4$. Both phase kinks are $\sim \pi$ at $t=T/4$ as shown in Fig.~\ref{fig:double}(c). When $t$ exceeds $T/4$ (Fig.~\ref{fig:double}(d)), the phase jumps by $2\pi$ in the two regions $r\lesssim L/4$ and $r \gtrsim 3L/4$ so that the winding number changes to $n=-1$ by losing the phase of $4\pi$ in total. It is worth stressing that this ``double phase slip" occurs without passing through a state with $n=0$. 
We note that there is no direct connection between the phase slip in real time and that in imaginary time~\cite{freire} (see also Supplementary Information, Section III). 
The dynamics in real time reflects the behavior of the average phase difference $\varphi(r,t)$, which comprises phase slips occurring at different times in different sites. 
This phase slip can be extracted from the superposition of two macroscopically distinct states with different winding numbers. 
At the same time the phase slip in imaginary time develops ``instantaneously" during underbarrier 
tunneling in contrast to the phase kink in real time that exhibits a sinusoidal oscillation and 
develops gradually. 
Nevertheless the similarity between the shape of the phase slip in Figs.~(\ref{fig:single}) and (\ref{fig:double}) and the expected shape of the kink in the instanton solution is very appealing.

\begin{figure}[t]
\includegraphics[scale=0.45]{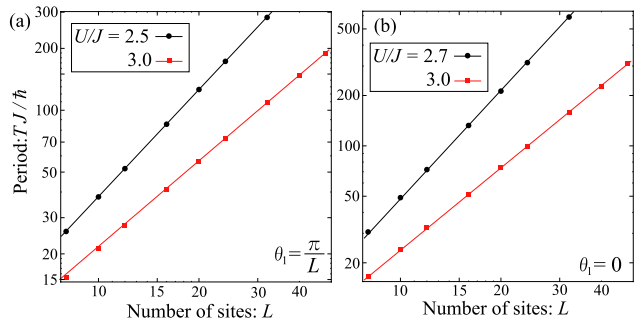}
\caption{\label{fig:tauL}
Period $T$ versus the number of sites $L$ for $\theta_1=\pi/L$ (a) and $\theta_1=0$ (b).
The filling factor is fixed to be $\nu = 1$. The plots are on a log-log scale.
}
\end{figure}
In the above calculations, we took a relatively small system size $L=16$ and unit filling ($N=L$).
Increasing the number of sites up to $L=48$, we checked that the basic properties
of the supercurrent dynamics mentioned above do not change. In Fig.~\ref{fig:tauL}, we show the period $T$ of the coherent oscillations as a function of $L$ on a log-log scale. There we clearly see that the period monotonically increases with $L$ following a power law, $T \propto L^{\alpha} $. Since the current $I$ at a fixed winding number is inversely proportional to $L$ this implies that the frequency of oscillations scales as a power of the current. This effect is similar to the situation happening in 2D superconductors at finite temperatures~\cite{ambegaokar}, where the supercurrent dissipation rate coming from vortex unbinding also scales as a power of the current. We also note that the commensurability of the filling factor is crucial for the coherent supercurrent dynamics. Only in case of commensurate fillings, the two states $|\Phi_1\rangle$ and $|\Phi_0\rangle$
(or $|\Phi_{-1}\rangle$) are coupled through the Umklapp scattering process~\cite{hallwood,ana}
and the coherent oscillations can occur.

\section{Comparison with the instanton method}
\begin{figure}[tb]
\includegraphics[scale=0.7]{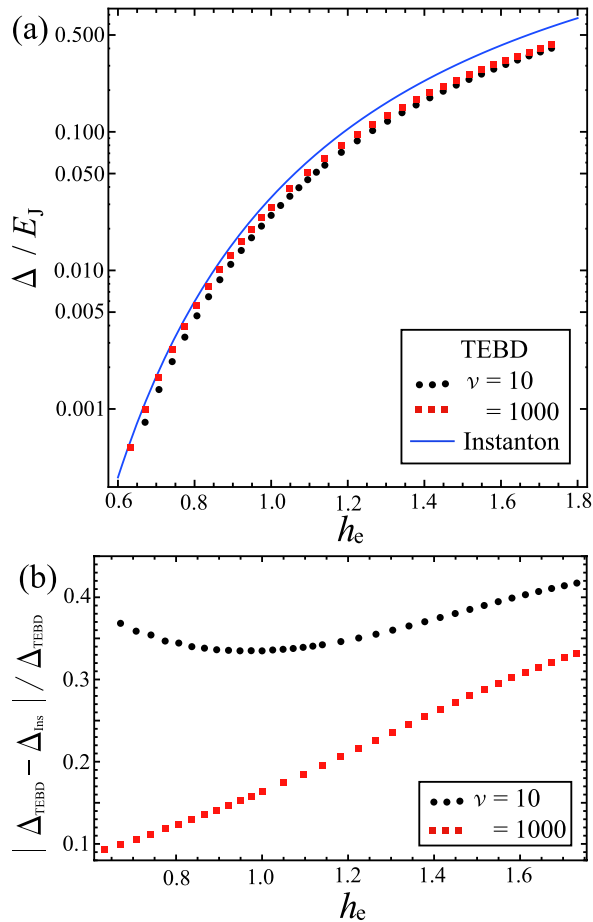}
\caption{\label{fig:comp}
(a) Energy splitting $\Delta/E_{\rm J}$ as a function of the effective Planck's constant
$h_{\rm e}\equiv \sqrt{U/(\nu J)}$ for $L=8$.
The blue solid line represents the result by the instanton method corresponding to
Eq.~(\ref{eq:espIns1}) with $\tilde{s}_I = 7.363$ and $K=3.06$.
The red squares and the black circles are the TEBD results for $\nu = 1000$ and $\nu =10$,
respectively.
(b) Ratio $|\Delta_{\rm TEBD} - \Delta_{\rm Ins}|/\Delta_{\rm TEBD}$ of the difference between
the TEBD and instanton results as a function of $h_{\rm e}$.
}
\end{figure}
Having established that the supercurrent dynamics exhibits coherent oscillations caused by MQT, 
we now compare the frequency of oscillations calculated by the instanton method with the TEBD 
results. 
For this purpose, we choose the situation of $\theta_1 = \pi/L$, characterized by a single phase slip 
dynamics. 
To seek a simple analytical expression of the energy splitting between the two current states 
$\Delta$, which is the same as the oscillation frequency, we assume $U \nu \gg J$ and $\nu \gg 1$. 
In this case the Bose-Hubbard model can be mapped onto the $O(2)$ quantum rotor model~\cite{anatoli2,sachdev}. 
This limit also describes a regular array of coupled Josephson junctions. 
We have confirmed that the quantum rotor model indeed gives a very accurate value of $\Delta$ 
when the filling is large: $\nu \gtrsim 1000$ (see Supplementary Information, Section II). 
In the quantum rotor model the phase space variables are the superfluid phases on each site 
and the conjugate momenta, corresponding to the fluctuations of the number of particles. 
So overall the phase space consists of $2L$ variables. Applying the instanton method to the 
quantum rotor model, we obtain the energy splitting expressed as
\begin{eqnarray}
\frac{\Delta}{E_{\rm J}}  = 2 L K  \sqrt{\frac{\tilde{s}_{I}}{2\pi h_{\rm e}}}
                             \exp\left(-\frac{\tilde{s}_I}{h_{\rm e}}\right),
\label{eq:espIns1}
\end{eqnarray}
where $E_{\rm J}\equiv \sqrt{\nu J U}$ is the Josephson plasma energy and
$h_{\rm e}\equiv \sqrt{U/(\nu J)}$ is the effective Planck's constant. Note that $h_{\rm e}\ll 1$ deep in the superfluid regime where number and phase can be approximately treated as classical variables. At $h_{\rm e}\sim 1$ the quantum fluctuations become important and can even drive the system to a different insulating phase~\cite{sachdev}. Since the instanton method is a generalization of the WKB semiclassical approximation~\cite{coleman1, vainstein}, the expression for the energy splitting (\ref{eq:espIns1}) is supposed to be accurate when $h_{\rm e}/\tilde{s}_I$ is sufficiently small. We note that the quantum rotor model has a clear advantage over original Bose-Hubbard model that the instanton action $\tilde{s}_I$ and the coefficient $K$ do not depend on $U/J$ and $\nu$ (see Methods and Supplementary Information for specific expressions of $\tilde{s}_I$ and $K$). In other words, the dependence of $\Delta/E_{\rm J}$ on $U/J$ and $\nu$ comes only through a single parameter $h_{\rm e}$. Using explicit calculation (see Methods) we obtain $\tilde{s}_I = 7.363$ and $K=3.06$ if we set the number of sites to be $L=8$.

We can also extract the energy splitting $\Delta$ from TEBD simulations by fitting the the overlap 
$f(t)=|\langle \Phi_1|\Psi(t)\rangle|^2$ (like in Fig.~\ref{fig:CON}(c)) using the function
\begin{eqnarray}
f(t)=B \cos^2\left(\frac{\Delta }{2\hbar}t \right) + C,
\end{eqnarray}
where $\Delta$, $B$, and $C$ are the free parameters. In Fig.~\ref{fig:comp}(a), we show the energy splitting $\Delta$ versus $h_{\rm e}$ calculated by the instanton method (blue solid line), and by TEBD for the filling factors $\nu = 1000$ (red squares) and $\nu = 10$ (black circles). It is evident that for $\nu = 1000$ and $h_{\rm e}$ sufficiently small the instanton and TEBD results agree very well. To quantify the error of the instanton method, in Fig.~\ref{fig:comp}(b) we show the relative difference between the two results: $|\Delta_{\rm TEBD} - \Delta_{\rm Ins}|/\Delta_{\rm TEBD}$. For $\nu = 1000$ (red squares), as $h_{\rm e}$ decreases, the error also decreases such that
it is within $10\%$ when $h_{\rm e}\lesssim 0.7$. It is hard to push the calculation to even smaller values of $h_{\rm e}$ because of exponential sensitivity of the period of oscillations to the effective Planck's constant. Nevertheless our results allow us to make the conclusion that the instanton method can provide quantitatively accurate prediction for the tunneling probability when $h_e/\tilde{s}_I$ is sufficiently small. At the same time, the error for $\nu = 10$ is significantly larger than that for $\nu = 1000$. Moreover the error does not even monotonically depend on $h_{\rm e}$. This clearly means that at this filling the quantum rotor model gives only qualitative description of the tunneling process.

\section{Summary}
We analyzed quantum dynamics of supercurrents of one-dimensional lattice bosons in a ring.
In particular, our focus was on the coherent oscillations between the two degenerate
current states via macroscopic quantum tunneling (MQT). 
The period of these oscillations $T$ is related to the energy splitting $\Delta$ induced by the 
tunneling as $T=2\pi\hbar/\Delta$. 
We calculated $\Delta$ both simulating real-time dynamics using the time-evolving block 
decimation (TEBD) method and within the imaginary time instanton method. 
We showed that the result of instanton calculation is in very good quantitative agreement with 
the TEBD result when the effective Planck's constant $h_{\rm e}$ is sufficiently small. 
This agreement verifies the instanton method applied to coherent MQT involving many 
collective variables.

We also want to emphasize that the success in applying TEBD (or equivalently the time-dependent
density matrix renormalization group) to MQT problems opens up new possibilities to analyzing macroscopic tunneling phenomena. In particular, (i) TEBD allows us to precisely calculate the
energy splitting even for large $h_{\rm e}$, where the instanton method fails. (ii) TEBD provides time evolution of the many-body wave function, from which one can calculate various quantities, for example different correlation functions.
The first advantage (i) is crucial for quantitative simulation of experiments (e.g. in cold gases), where it is easier to work in the regime of larger $h_{\rm e}$ and shorter periods to avoid various effects of decoherence like particle losses. Moreover, the second advantage allows one to reveal detailed processes of MQT in real time. As an example, we have analyzed the time evolution of the phase-phase correlation functions and revealed the existence of the phase slips associated with the coherent oscillations, which can be detected in experiments. One can extend this analysis to study higher order correlation functions to e.g. detect shot noise of phase slips or even their full counting statistics~\cite{levitov}.

\begin{acknowledgments}
I. D. thanks M. Nishida, A. Nunnenkamp, T. Nikuni, S. Kurihara, and Y. Kato for valuable comments
and discussions. I. D. acknowledges support from a Grant-in-Aid from JSPS. I.D. is grateful to Boston University visitors program for hospitality. A. P. was supported by AFOSR YIP and Sloan Foundation.
\end{acknowledgments}
\section{Methods}

Here, we outline instanton derivation of the energy splitting~(\ref{eq:espIns1}) and give specific expressions of $\tilde{s}_I$ and $K$. For additional details of the derivation of these expressions we refer the reader to the Supplementary Information, Section III. In the limit when $U\nu \gg J$ and $\nu \gg 1$, number fluctuations are significantly suppressed and the Bose-Habbard model Eq.~(\ref{eq:BHH}) can be mapped onto the $O(2)$ quantum rotor
model~\cite{anatoli2}, described by the effective action
\begin{eqnarray}
\tilde{s} = \int_{-\frac{\tilde{\beta}}{2}}^{\frac{\tilde{\beta}}{2}}
d\tilde{\tau} \left[
\frac{1}{2}\frac{\partial \vec{\phi}}{\partial \tilde{\tau}}\cdot \frac{\partial \vec{\phi}}{\partial \tilde{\tau}}
+V(\vec{\phi})
\right].
\label{eq:sVecM}
\end{eqnarray}
where $\vec{\phi}$ is an $L$-dimensional vector defined as
\begin{eqnarray}
\vec{\phi} =
\left(
\phi_{1}(\tilde{\tau}),\ldots, \phi_{j}(\tilde{\tau}), \ldots, \phi_{L}(\tilde{\tau})
\right)^{\bf t}
\end{eqnarray}
and the potential is
\begin{eqnarray}
V(\vec{\phi}) &=& \sum_{j=1}^{L} V_j(\phi_{j+1},\phi_j) \nonumber\\
&=& \sum_{j=1}^{L}  -2\cos\left( \phi_{j+1} - \phi_j - \theta \right).
\end{eqnarray}
$\phi_j$ is the phase of particles at the $j$-th site, and $\tilde{\tau}\equiv \tau E_{\rm J}/\hbar$
and $\tilde{\beta}\equiv E_{\rm J}/(k_{\rm B} T)$ denote the imaginary time and the inverse
temperature in the Josephson plasma energy unit.
We remind that $E_{\rm J}\equiv \sqrt{\nu J U}$.

Extremizing the action, by setting $\delta \tilde{s} = 0$, we obtain the classical equations of motion
for the phases:
\begin{eqnarray}
\frac{\partial^2 \phi_j}{\partial \tilde{\tau}^2} =
2\sin\left(\phi_{j+1}-\phi_j -\theta \right)
- 2\sin\left(\phi_{j}-\phi_{j-1} -\theta \right).
\label{eq:classical1M}
\end{eqnarray}
In order to calculate $\tilde{s}_I$, we numerically find the instanton solution
$\vec{\phi}(\tilde{\tau})=\vec{\phi}_I(\tilde{\tau})$ of Eq.~(\ref{eq:classical1M}), which connects the two degenerate states with different winding numbers:
\begin{eqnarray}
\phi_j(-\tilde{\beta}/2) = \frac{2\pi j}{L}-\pi\left(1+\frac{1}{L}\right),
\,\,\,
\phi_j(\tilde{\beta}/2) = 0.
\label{eq:insBoundM}
\end{eqnarray}
(see Fig.~7 of the Supplementary Information). Once this solution is obtained, we obtain $\tilde{s}_I$ by substituting
$\vec{\phi}(\tilde{\tau})=\vec{\phi}_I(\tilde{\tau})$ into Eq.~(\ref{eq:sVecM}):
\begin{eqnarray}
\tilde{s}_I = \int_{-\frac{\tilde{\beta}}{2}}^{\frac{\tilde{\beta}}{2}}
d\tilde{\tau} \left[
\frac{1}{2}\frac{\partial \vec{\phi}_I}{\partial \tilde{\tau}}\cdot \frac{\partial \vec{\phi}_I}{\partial \tilde{\tau}}
+V(\vec{\phi}_I)
\right].
\label{eq:sInsM}
\end{eqnarray}
For $L=8$, Eq.~(\ref{eq:sInsM}) gives $\tilde{s}_I = 7.363$.

In turn the prefactor $K$ is given by~\cite{coleman2, vainstein}
\begin{eqnarray}
K=\left(\frac{\prod_{m}\lambda_m^{(0)}}{\prod_{m \neq 0} \lambda_m}\right)^{1/2},
\label{eq:KM}
\end{eqnarray}
where $\lambda_m$'s and $\lambda_m^{(0)}$'s are the solutions of the following eigenvalue equations:
\begin{eqnarray}
\hat{{\cal M}}\vec{\xi}_m(\tilde{\tau})=\lambda_m\vec{\xi}_m(\tilde{\tau}),
\label{eq:eigEqM1M}
\end{eqnarray}
and
\begin{eqnarray}
\hat{{\cal M}}^{(0)}\vec{\xi}_m^{(0)}(\tilde{\tau})=\lambda_m^{(0)}\vec{\xi}_m^{(0)}(\tilde{\tau}).
\label{eq:eigEqM2M}
\end{eqnarray}
Here the $L$-dimensional vectors
\begin{eqnarray}
\vec{\xi}_m =
\left(
\xi_{1,m}(\tilde{\tau}),\ldots, \xi_{j,m}(\tilde{\tau}), \ldots, \xi_{L,m}(\tilde{\tau})
\right)^{\bf t},
\end{eqnarray}
and
\begin{eqnarray}
\vec{\xi}_m^{(0)} =
\left(
\xi_{1,m}^{(0)}(\tilde{\tau}),\ldots, \xi_{j,m}^{(0)}(\tilde{\tau}), \ldots, \xi_{L,m}^{(0)}(\tilde{\tau})
\right)^{\bf t}.
\end{eqnarray}
obey the orthonormalization condition
\begin{eqnarray}
\int d\tilde{\tau} \,\, \vec{\xi}_l \cdot \vec{\xi}_m = \delta_{l,m},
\,\,\,
\int d\tilde{\tau} \,\, \vec{\xi}_l^{(0)} \cdot \vec{\xi}_m^{(0)} = \delta_{l,m}.
\end{eqnarray}
The $L\times L$ matrices $\hat{{\cal M}}$ and $\hat{{\cal M}}^{(0)}$ are defined as
\begin{eqnarray}
{\cal M}_{j,k}&&\!\!\!\!=\delta_{j,k}
\left(
-\frac{\partial^2}{\partial \tau^2}
+ \left. \frac{\partial^2 V_j}{\partial \phi_j^2} \right|_{\vec{\phi}=\vec{\phi}^I}
+ \left. \frac{\partial^2 V_{j-1}}{\partial \phi_j^2} \right|_{\vec{\phi}=\vec{\phi}^I}
\right)
\nonumber\\
&&\!\!\!\!\!\!\!\!\!\!\!\!\!\!\!\!\!\!\!\!\! + \delta_{j, k-1}
\left. \frac{\partial^2 V_j}{\partial \phi_j \partial \phi_{j+1}} \right|_{\vec{\phi}=\vec{\phi}^I}
+ \delta_{j, k+1}
\left. \frac{\partial^2 V_{j-1}}{\partial \phi_j \partial \phi_{j-1}} \right|_{\vec{\phi}=\vec{\phi}^I},
\end{eqnarray}
and
\begin{eqnarray}
{\cal M}_{j,k}^{(0)}=\delta_{j,k}
\left(
-\frac{\partial^2}{\partial \tau^2}
+2 \omega^2
\right)
- \delta_{j, k-1} \omega^2
- \delta_{j, k+1} \omega^2,
\end{eqnarray}
where $\omega^2= \left. \partial_{\phi_j}^2 V_j \right|_{\vec{\phi}=\vec{0}}$.
Notice that $(L+1)$-th and $0$th sites are equivalent to 1st and $L$-th sites, respectively,
reflecting the periodic boundary condition. For $L=8$, Eq.~(\ref{eq:KM}) gives $K = 3.06$.


\section{Supplementary Information}
\section{I. TEBD for large filling factors}
\begin{figure}[b]
\includegraphics[scale=0.7]{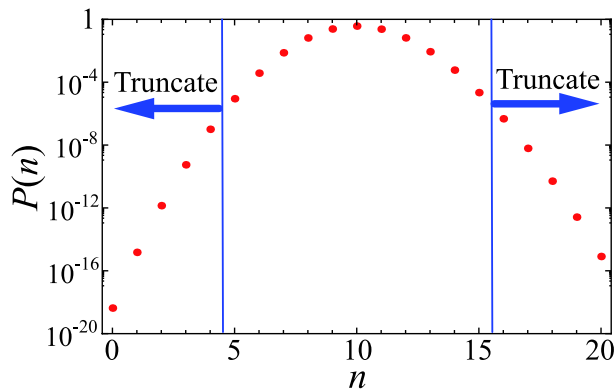}
\caption{\label{fig:OP}
Occupation probability $P(n)$ (in the log-scale) of the local Fock state $|n\rangle$ in the ground state of the untwisted Bose-Hubbard model with $L=8$, $\nu = 10$, and $U/J = 5$. Here $L$ is the system size, $\nu$ is the filling factor, $U$ is the onsite interaction and $J$ is the hopping energy (see Eq.~(1) in the main text).}
\end{figure}
In this section, we present an idea of adopting the time-evolving block decimation (TEBD) method to the Bose-Hubbard model when the average number of particles per site $\nu$ (or the filling factor) is large. The key of the idea is that in addition to the upper bound, the lower bound for the occupation number of particle per site is introduced in order to significantly reduce the size of the local Hilbert space. This idea is crucial because the quantitative comparison of the TEBD results with the results of the instanton method based on the quantum rotor model is possible only for very large $\nu \gtrsim 1000$ (see Sec.~II).

Let us consider a system described by the one-dimensional (1D) Bose-Hubbard model with
$L$ lattice sites.
Spanning the Hilbert space of the whole system by a product of local Hibert spaces
of dimension $d$, a many-body wave function of the system is expressed as
\begin{eqnarray}
|\Psi\rangle =
\sum_{j_1,j_2,\ldots,j_L=1}^{d} c_{j_1,j_2,\ldots,j_L}
|j_1\rangle |j_2\rangle \cdots |j_L\rangle.
\end{eqnarray}
In the TEBD algorithm~\cite{vidal1}, coefficients $c_{j_1,j_2,\ldots,j_L}$ are decomposed
in a particular matrix product form as
\begin{eqnarray}
c_{j_1,j_2,\ldots,j_L}
\!\!\!&=&\!\!\!
\sum_{\alpha_1,\ldots,\alpha_{L-1}=1}^{\chi}\!\!\!
\Gamma_{\alpha_1}^{[1]j_1}\lambda_{\alpha_1}^{[1]}
\Gamma_{\alpha_1\alpha_2}^{[2]j_2}\lambda_{\alpha_2}^{[2]} \cdots
\lambda_{\alpha_{L-2}}^{[L-2]}
\Gamma_{\alpha_{L-2}\alpha_{L-1}}^{[L-1]j_{L-1}}
\lambda_{\alpha_{L-1}}^{[L-1]}\Gamma_{\alpha_{L-1}}^{[L]j_L}.
\label{eq:tensorproduct}
\end{eqnarray}
The vector $\lambda_{\alpha_l}^{[l]}$ represents the coefficients of the
Schmidt decomposition of $|\Psi\rangle$ with respect to the bipartite splitting
of the system into $[1,\ldots,l-1,l]:[l+1,l+2,\ldots,L]$. 
The tensors $\Gamma$'s constitute the Schmidt vectors together with the $\lambda$-vectors.
$\chi$ is the number of basis states, which is taken to be sufficiently large so that the
error due to this truncation is nearly equal to zero.
In the typical calculations in the main text, it ranges from $\chi=100$ to $\chi=200$.

Usually dimension of the local Hilbert space corresponding to a single site is chosen as $d=n_{\rm max}+1$, where
$n_{\rm max}$ is the maximum number of particles per site. It is spanned by the basis set, $\left\{|n=0\rangle, |1\rangle, \ldots, |n_{\rm max}-1\rangle, |n_{\rm max}\rangle \right\}$. While in principle $n_{\rm max}$ is equal to the total number of particles in the system, taking much smaller $n_{\rm max}$  provides converged results in practice.
For instance, for accurate determination of the zero temperature phase diagram of the Bose Hubbard model at unit-filling, $n_{\rm max} = 5$ ($d=6$) is sufficient~\cite{kuehner}. At large filling factors, however, this choice of the local Hilbert space basis makes computations extremely expensive, because the computational cost in TEBD scales
as $L d^3 \chi^3$~\cite{vidal1}. To solve this problem, in addition to $n_{\rm max}$, we introduce the minimum number of particles per site $n_{\rm min}$ and span the local Hilbert space by the basis set, $\left\{|n=n_{\rm min}\rangle, |n_{\rm min}+1\rangle, \ldots, |n_{\rm max}-1\rangle, |n_{\rm max}\rangle \right\}$, and thus $d=n_{\rm max}-n_{\rm min}+1$. In the parameter region of $U/(\nu J)\sim 1$, where our TEBD simulations are carried out,
setting $n_{\max}=\nu+5$ and $n_{\min}=\nu-5$ corresponding to $d=11$ is sufficient for the convergence
regardless of the value of $\nu$. For instance, in Fig.~\ref{fig:OP}, we plot the occupation probability $P(n)$ of the local Fock state $|n\rangle$ for $L=8$, $\nu = 10$, and $U/J = 5$. It is evident that $P(n)$ exponentially decays as $n$ deviates from its average $\nu = 10$ and that $P(n)$ for $n>15$ and $n<6$ is less than $10^{-6}$. This justifies this truncation scheme for practical calculations.

\section{II. Effective action for the phase slip problem}
\label{sec:QR}
In this section, we explain the mapping of the Bose-Hubbard model onto the $O(2)$-quantum
rotor model.
For this purpose, we start with the grand canonical partition function,
\begin{eqnarray}
Z=\int {\cal D}b^{\ast}{\cal D}b \exp\left\{-\frac{S[b^{\ast},b]}{\hbar}\right\}
\label{part_func}
\end{eqnarray}
where the action $S[b^{\ast},b]$ for the Bose-Hubbard model (see Eq.~(1) in the main text) is given by
\begin{eqnarray}
S[b^{\ast},b]=\sum_{j=1}^L \int_{-\frac{\hbar\beta}{2}}^{\frac{\hbar\beta}{2}} d\tau &&
\left[
b_j^{\ast}(\tau)\hbar\frac{\partial}{\partial\tau}b_j(\tau)
- J\left( b_{j}^{\ast}(\tau)b_{j+1}(\tau) e^{-i\theta}+ b_{j+1}^{\ast}(\tau)b_{j}(\tau) e^{i\theta}\right)
\right.
\nonumber\\
&&\left.
+ \frac{U}{2}b_j^{\ast}(\tau)b_j^{\ast}(\tau)b_{j}(\tau)b_{j}(\tau)-\mu b_j^{\ast}(\tau) b_j(\tau)
\right]
\end{eqnarray}
where $U$ is the onsite interaction, $J$ is the hopping energy, $\nu$ is the filling factor, $\mu\approx U\nu$ is the chemical potential, and $\theta$ is the phase twist. For convenience we introduce finite small temperature $T$ corresponding to the inverse temperature $\beta \equiv (k_B T)^{-1}$. In the end of calculations we will take the limit $T\to 0$. Inserting $b_j=\sqrt{n_j}e^{i\phi_j}$, the action is rewritten as
\begin{eqnarray}
S[n,\phi]=\sum_{j=1}^L \int_{-\frac{\hbar\beta}{2}}^{\frac{\hbar\beta}{2}} d\tau
\left[
\hbar n_j\left( i \frac{\partial \phi_j}{\partial \tau} +\frac{1}{2n_j}\frac{\partial n_j}{\partial \tau}\right)
-2\sqrt{n_j n_{j+1}}J\cos\left(\phi_{j+1}-\phi_j-\theta\right)
+\frac{U}{2}(n_j-\nu)^2
\right]
\end{eqnarray}
Splitting the number of particles per site into its average and fluctuation as $n_j = \nu + \delta n_j$, assuming that $\nu$ is integer and that $U\nu \gg J$ and $\nu \gg \delta n_j$, we find that the action is then approximated as
\begin{eqnarray}
S[n,\phi]=\sum_{j=1}^L \int_{-\frac{\hbar\beta}{2}}^{\frac{\hbar\beta}{2}} d\tau
\left[
i\hbar\, \delta n_j \frac{\partial \phi_j}{\partial \tau}
-2\nu J\cos\left(\phi_{j+1}-\phi_j-\theta\right)
+\frac{U}{2}\delta n_j^2
\right]
\label{eq:actionNphi}
\end{eqnarray}
Since Eq.~(\ref{eq:actionNphi}) contains only the linear and quadratic terms with respect to number fluctuations $\delta n_j$,
these degrees of freedom can be integrated out. Then, the action is described in terms of the phases as
\begin{eqnarray}
S[\phi]=\sum_{j=1}^L \int_{-\frac{\hbar\beta}{2}}^{\frac{\hbar\beta}{2}} d\tau
\left[
\frac{\hbar^2}{2U}  \left(\frac{\partial \phi_j}{\partial \tau}\right)^2
-2\nu J\cos\left(\phi_{j+1}-\phi_j-\theta\right)
\right].
\label{ac_phi}
\end{eqnarray}
%
It is convenient to change variables
\begin{eqnarray}
\tau = \frac{\hbar}{\sqrt{\nu J U}} \tilde{\tau},
\end{eqnarray}
so that Eq.~(\ref{ac_phi}) is rewritten as
\begin{eqnarray}
S=\hbar \sqrt{\frac{\nu J}{U}}\tilde{s},
\label{eq:heff}
\end{eqnarray}
where $\tilde{s}$ is the dimensionless action
\begin{eqnarray}
\tilde{s}[\phi] = \sum_{j=1}^{L}\int_{-\tilde{\beta}/2}^{\tilde{\beta}/2}
d\tilde{\tau}
\left[
\frac{1}{2}\left(\frac{d\phi_j}{d\tilde{\tau}}\right)^2
-2\cos(\phi_{j+1}-\phi_j-\theta)
\right].
\label{eq:sDless}
\end{eqnarray}
and $\tilde{\beta} = \beta \sqrt{\nu J U}$. From Eqs.~(\ref{part_func}) and (\ref{eq:heff}) we clearly see that $h_{\rm e}\equiv \sqrt{U/(\nu J)}$ plays the role of the effective dimensionless Planck's constant for this problem with $h_{\rm e}\to 0$ corresponding to the classical (Bogoliubov) limit and $h_{\rm e}\gtrsim 1$ corresponding to the regime of strong quantum fluctuations.

Extremizing the action by imposing $\delta \tilde{s} = 0$, we obtain the classical equations of motion
for the phases $\phi_j$,
\begin{eqnarray}
\frac{\partial^2 \phi_j}{\partial \tilde{\tau}^2} =
2\sin\left(\phi_{j+1}-\phi_j -\theta \right)
- 2\sin\left(\phi_{j}-\phi_{j-1} -\theta \right).
\label{eq:classical}
\end{eqnarray}
There are two types of stationary solution of Eq.~(\ref{eq:classical}).
One is
\begin{eqnarray}
\phi_j = \frac{2\pi n}{L}(j-1),
\label{eq:meta}
\end{eqnarray}
which describes the current carrying states with the winding-number $n$. The other is a saddle-point solution with a phase kink separating (meta)stable states with different winding numbers:
\begin{eqnarray}
\phi_j =\frac{\alpha}{2}+\varphi(j-1),
\end{eqnarray}
where
\begin{eqnarray}
\alpha = - \pi\frac{L-1+2n}{L-2}+2\theta\frac{L-1}{L-2}\mod 2\pi,
\label{eq:saddle}
\end{eqnarray}
and
\begin{eqnarray}
\varphi = \frac{2\pi n-\alpha}{L-1}.
\end{eqnarray}
Notice that in Eq.~(\ref{eq:saddle}) the phase kink is assumed to be located at the link
between the 1st and $L$-th sites. The magnitude of this kink $\alpha$ is defined within the interval $[-2\pi,0]$. In the limit of the large number of sites $L\gg 1$ the expression for $\alpha$ simplifies:
\begin{equation}
\alpha\approx -\pi\left(1+{2 n\over L}\right)+2\theta\mod 2\pi.
\end{equation}
In particular, in the case $n=0$ and $\theta = \pi / L$, which we are interested in, $\alpha = -\pi(1-1/L)\approx -\pi$ and $\varphi = \pi /L$.

For $\theta = \pi /L$, the two current states with windings $n=1$ and $n=0$ are degenerate. Quantum tunneling couples them and breaks the degeneracy, leading to the energy splitting $\Delta$ between the ground (bonding) state and the first-excited (anti-bonding) state. This tunneling process is associated with generation of a ``phase slip" or equivalently a phase kink. In imaginary time evolution the virtual kink forms during the imaginary time evolution of the phase between the two current states. If the state of Eq.~(\ref{eq:meta}) with $n=1$ is prepared initially, the many-body wave function coherently oscillates with the period $2\pi \hbar/\Delta$ between the states with $n=1$ and $n=0$. It is well known~\cite{sakita} that the energy splitting can be expressed as
\begin{eqnarray}
\Delta = 2\lim_{\beta \rightarrow \infty} \frac{A}{\beta},
\label{eq:espOrig}
\end{eqnarray}
where
\begin{eqnarray}
A \equiv \frac{Z_1}{Z_0}
\label{eq:capia}
\end{eqnarray}
and
\begin{eqnarray}
Z_1=\int_{(1)}{\cal D}\phi \exp\left\{-\frac{\tilde{s}[\phi]}{h_{\rm e}}\right\},\quad
Z_0=\int_{(0)}{\cal D}\phi  \exp\left\{-\frac{\tilde{s}[\phi]}{h_{\rm e}}\right\}.
\end{eqnarray}
Notice that $\int_{(1)}{\cal D}\phi$ denotes the path integral over trajectories
containing a single instanton, while $\int_{(0)}{\cal D}\phi$ is the path integral containing zero
instantons. According to the instanton method~\cite{coleman1,coleman2}, the energy splitting of
Eq.~(\ref{eq:espOrig}) is well approximated by
\begin{eqnarray}
\Delta  \simeq 2 L K E_{\rm J} \sqrt{\frac{\tilde{s}_{I}}{2\pi h_{\rm e}}}
                             \exp\left(-\frac{\tilde{s}_I}{h_{\rm e}}\right),
\label{eq:espIns}
\end{eqnarray}
where $\tilde{s}_I$ denotes the action for the instanton solution, $K$ is the constant we define below, and
$E_{\rm J}\equiv\sqrt{\nu J U}$ is the Josephson plasma energy. It is worth stressing that $\tilde{s}_I$ and $K$ do not depend on $\nu$ and $U/J$, but depend only on $L$ so that $\Delta / E_{\rm J}$ depends on $U, J$, and $\nu$ only through $h_{\rm e}$. In the following section we will present a derivation of Eq.~(\ref{eq:espIns}) first evaluating it approximately and then exactly and will give the explicit form of the coefficient $K$.

\begin{figure}[tb]
\includegraphics[scale=0.6]{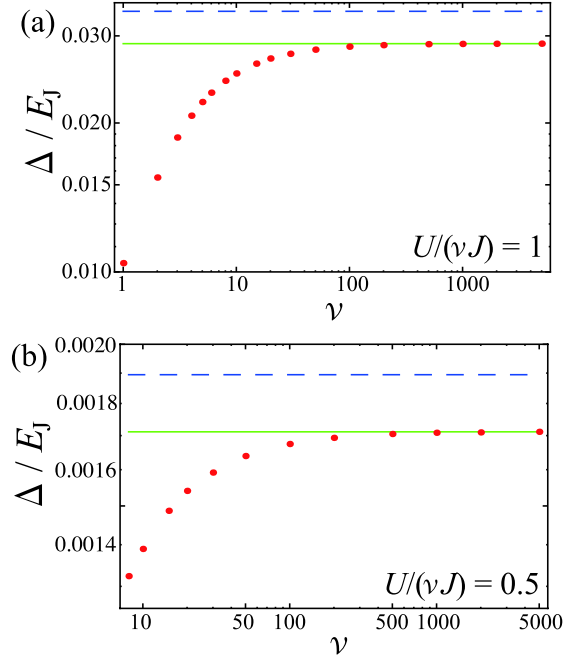}
\caption{\label{fig:Lnu}
Ratio of the energy splitting $\Delta$ to the Josephson energy $E_J=\sqrt{\nu JU}$ as a 
function of $\nu$ for $U/(\nu J)\equiv h_{\rm e}^2 = 1$ (a) and $U/(\nu J) = 0.5$ (b). 
The solid lines represent the value of $\Delta/E_{\rm J}$ at $\nu = 5000$. 
The dashed lines are the prediction of the instanton method.
Notice that both $U$ and $\nu$ are changed for a fixed value of $J$ such that 
$h_{\rm e}$ remains constant.
}
\end{figure}
As mentioned above, the mapping onto the quantum rotor model is justified when
$U\nu\gg J$ and $\nu \gg 1$.
For unambiguous comparison between the TEBD and instanton results, we need to
specify quantitatively the parameter region where the quantum rotor model is valid
for the calculations of the energy splitting. It is clear in Eq.~(\ref{eq:espIns}) as $\nu$ increases at a fixed value of $h_{\rm e}$, $\Delta/E_{\rm J}$ should saturate at a constant corresponding to the quantum rotor limit. As we show in Fig.~\ref{fig:Lnu}, where we plot $\Delta/E_{\rm J}$ versus $\nu$ for two different values of $h_{\rm e}$, this is indeed the case. Note that this ratio $\Delta/E_{\rm J}$ becomes independent on $\nu$ only at very large filliwng factors $\nu\gtrsim 1000$. We also want to point that for the smaller value of $h_{\rm e}$ the larger the filling factor is required for the convergence. Since our quantitative analysis in the main text is focused on the region of $U/(\nu J) \gtrsim 0.5$, the quantum rotor model is sufficiently accurate for $\nu = 1000$, which we use in practice.

\section{III. Instanton method for the quantum rotor model}
\subsection{A. One collective variable}
\label{sec:one}
\begin{figure}[b]
\includegraphics[scale=0.7]{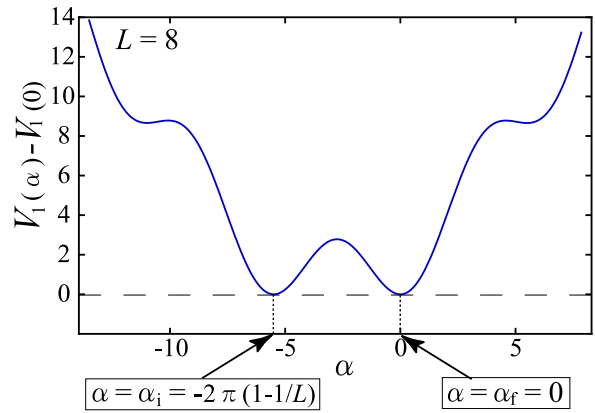}
\caption{\label{fig:1Dpote}
Effective potential $V_1(\alpha)$ for $\theta=\pi/L$ and $L=8$.
}
\end{figure}
\begin{figure}[t]
\includegraphics[scale=0.7]{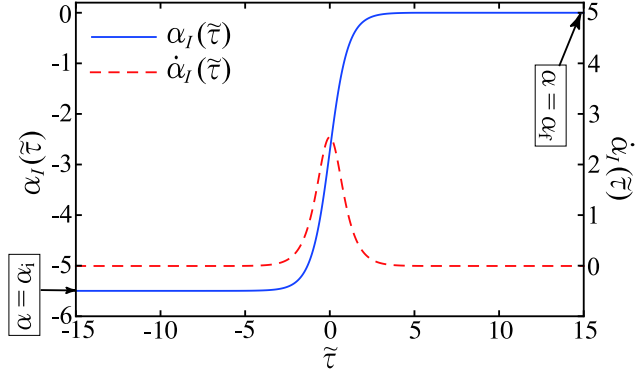}
\caption{\label{fig:ins1D}
Instanton solution $\alpha_I(\tilde{\tau})$ and
$\partial_{\tilde{\tau}}\alpha_I(\tilde{\tau})$.
}
\end{figure}
Since the instanton method in the presence of many degrees of freedom is quite complicated, 
we first use a simpler model, which is reduced from the quantum rotor model by assuming that 
the phase slip is described by only a single collective variable. 
This simple model, which represents a variational estimate of the full result, is useful to understand 
basic ideas of the calculation. 
Later we will generalize the result to the phase slip described by two degrees of freedom and 
finally show the complete instanton solution of the full problem.  
In the regime of validity of the quantum rotor model, the healing length $\xi =d \sqrt{2J/(\nu U)}$ is 
much shorter than the lattice spacing $d$. 
Hence the phase slip that develops during the tunneling process should be localized with in a few sites. Without loss of generality, we can assume that the phase kink develops at the link between the 1st 
and $L$-th sites. 
In the first approximation we assume that the phases along the instanton trajectory satisfy the ansatz,
\begin{eqnarray}
\phi_j(\tilde{\tau}) = \frac{\alpha(\tilde{\tau})}{2}+\varphi(\tilde{\tau}) (j-1),
\label{eq:intpl}
\end{eqnarray}
where $\alpha$ denotes the phase difference between the 1st and $L$-th sites. 
Its time dependence is found from extremizing the effective action (see below). 
The remaining phases on other sites are chosen as a simple linear function of the site index $j$ with 
$\varphi \equiv -\alpha/(L-1)$ such that the boundary condition $\phi_{L} = - \alpha/2$ is fulfilled. Substituting Eq.~(\ref{eq:intpl}) into Eq.~(\ref{eq:sDless}) we find that the effective dimensionless 
action describing the system becomes
\begin{eqnarray}
\tilde{s}[\alpha] = \int d\tilde{\tau} \left[
\frac{M}{2} \left(\frac{\partial \alpha}{\partial \tilde{\tau}}\right)^2
+ V_{1}(\alpha) \right].
\end{eqnarray}
This is nothing but the classical action of a particle with the effective mass $M$, which depends on the system size according to
\begin{eqnarray}
M=\frac{L(L+1)}{12(L-1)},
\end{eqnarray}
moving in the effective potential $-V_1(\alpha)$, where
\begin{eqnarray}
V_1(\alpha) = -2\cos\left(\alpha - \theta \right) -2(L-1)\cos(\varphi - \theta).
\end{eqnarray}
The shape of $V_1(\alpha) - V_1(0)$ for $\theta = \pi/L$ and $L=8$ is depicted
in Fig.~\ref{fig:1Dpote}. Note that $V_1(\alpha)$ has two global minima $\alpha = \alpha_i \equiv -2\pi(1-1/L)$ and $\alpha = \alpha_f \equiv 0$ corresponding to the current-carrying states with winding numbers $n=1$ and $n=0$ respectively. These two minima are separated by a local maximum, $\alpha = \alpha_s \equiv -\pi(1-1/L)$ describing
the saddle-point solution of Eq.~(\ref{eq:saddle}). Thus, introducing the collective variable $\alpha$, the phase slip problem is equivalent to tunneling of a single particle in a one-dimensional symmetric double-well potential. The corresponding classical equation of motion describing the particle motion in the (inverted) potential $-V_1(\alpha)$ is
\begin{eqnarray}
-M\frac{\partial^2\alpha}{\partial\tau^2}+\frac{\partial V_1}{\partial \alpha} =0.
\label{eq:1Deqm}
\end{eqnarray}
The instanton solution of this equation $\alpha(\tilde{\tau})=\alpha_{I}(\tilde{\tau})$ is the one satisfying the boundary conditions $\alpha(-\tilde{\beta}/2) = \alpha_i$ and  $\alpha(\tilde{\beta}/2) = \alpha_{f}$. Such a solution (shown in Fig.~\ref{fig:ins1D}) contains a kink in the phase $\alpha$. The instanton solution defines the classical trajectory in the path integral of $Z_1$. There is another trivial solution of Eq.~(\ref{eq:1Deqm}), $\alpha(\tilde{\tau}) = \alpha_i$ (or equivalently $\alpha(\tilde{\tau}) = \alpha_f$), which is the classical trajectory corresponding to the path integral of $Z_0$.

To calculate the ratio $A$ in Eq.~(\ref{eq:capia}) (see Ref.~\cite{vainstein} for more details), we substitute
\begin{eqnarray}
\alpha(\tilde{\tau})=\alpha_I(\tilde{\tau}) + \sqrt{\frac{h_{\rm e}}{M}}\sum_m c_m \xi_m(\tilde{\tau}),
\end{eqnarray}
into $Z_1$ and
\begin{eqnarray}
\alpha(\tilde{\tau})=\alpha_i + \sqrt{\frac{h_{\rm e}}{M}}\sum_m c_m \xi_m^{(0)}(\tilde{\tau}),
\end{eqnarray}
into $Z_0$, where $\xi_m$'s and $\xi_m^{(0)}$'s are complete sets of real orthonormal functions
obeying the following eigenvalue equations:
\begin{eqnarray}
\left(
-\frac{\partial^2}{\partial \tilde{\tau}^2}
+\frac{1}{M}\left. \frac{\partial^2 V_1}{\partial \alpha^2}\right|_{\alpha=\alpha_{\rm cl}}
\right)
\xi_m(\tilde{\tau}) = \lambda_m \xi_m(\tilde{\tau})
\label{eq:eigEq1}
\end{eqnarray}
and
\begin{eqnarray}
\left(
-\frac{\partial^2}{\partial \tilde{\tau}^2}
+\omega^2
\right)
\xi_m^{(0)}(\tilde{\tau}) = \lambda_m^{(0)} \xi_m^{(0)}(\tilde{\tau}),
\label{eq:eigEq2}
\end{eqnarray}
with $\omega^2=M^{-1}\partial_{\alpha}^2V_1|_{\alpha=\alpha_i}$.
Neglecting the terms higher than the second order with respect to $\sqrt{h_{\rm e}/M}$,
$A$ is approximated as
\begin{eqnarray}
A \simeq \exp\left(-\frac{\tilde{s}_I}{h_{\rm e}}\right)
\frac{\int \cdots \int \prod_m (2\pi)^{-1/2}dc_m\exp \left[-\frac{1}{2}\sum_m \lambda_m c_m^2\right]}
{\int \cdots \int \prod_m (2\pi)^{-1/2}dc_m\exp \left[-\frac{1}{2}\sum_m \lambda_m^{(0)} c_m^2\right]}
\label{eq:A2}
\end{eqnarray}
where $\tilde{s}_I$ is the action of the instanton solution given by
\begin{eqnarray}
\tilde{s}_I=\int d\tilde{\tau} M\left(\frac{\partial \alpha_{I}}{\partial \tilde{\tau}}\right)^2.
\end{eqnarray}
To carry out the integrals with respect to $c_m$'s in Eq~(\ref{eq:A2}), it is important that
due to the translation invariance of the instanton solution in the imaginary time,
Eq.~(\ref{eq:eigEq1}) possesses one solution $\xi_0$ with the eigenvalue $\lambda_0=0$.
For this zero mode, the integral in Eq.~(\ref{eq:A2}) is formally divergent.
To solve this problem, one needs simply replace $\int dc_0$ with $\sqrt{\tilde{s}_I/h_{\rm e}}\int d\tilde{\tau}$~\cite{coleman2} leading to
\begin{eqnarray}
A=\beta\sqrt{\nu J U}
\left(\frac{\prod_{m}\lambda_m^{(0)}}{\prod_{m \neq 0} \lambda_m}\right)^{1/2}
\sqrt{\frac{\tilde{s}_I}{2\pi h_{\rm e}}}
 \exp\left(-\frac{\tilde{s}_I}{h_{\rm e}}\right).
 \label{eq:A3}
\end{eqnarray}
In the above discussion, we assumed that the phase kink develops at the link between the 1st and 
$L$-th sites. 
In total there are  $L$ independent possibilities for the kink. 
Note that because we are dealing with a discrete system, there is no continuous symmetry 
associated with this degeneracy and thus no additional zero eigenvalue in Eq.~(\ref{eq:A3}). 
All instanton solutions centered around different links give identical contribution to $Z_1$. 
It is therefore only necessary to multiply $A$ by $L$ before substituting it into Eq.~(\ref{eq:espOrig}). Thus, we obtain Eq.~(\ref{eq:espIns}) with the coefficient
\begin{eqnarray}
K=\left(\frac{\prod_{m}\lambda_m^{(0)}}{\prod_{m \neq 0} \lambda_m}\right)^{1/2}.
\label{eq:K}
\end{eqnarray}
Now both $s_{I}$ and $K$ can be straightforwardly found numerically. For the situation of a single collective variable described here they are explicitly given in the first row of Table~\ref{tab:sK}.

\subsection{B. Two collective variables}
\begin{figure}[b]
\includegraphics[scale=0.5]{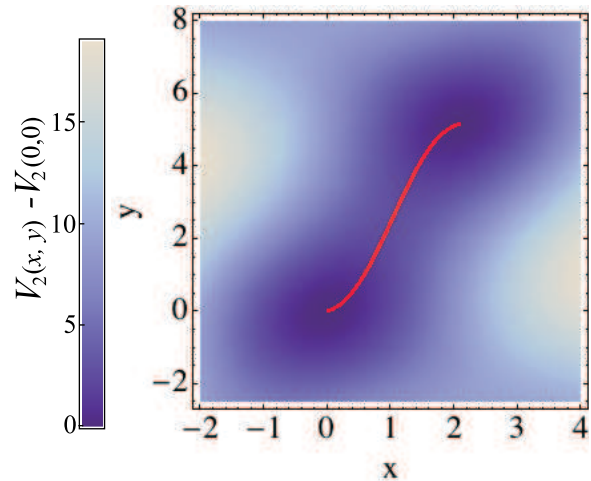}
\caption{\label{fig:2Dpote}
Effective potential $V_2(x,y)-V_2(0,0)$ for $\theta=\pi/L$ and $L=8$.
The solid line represent the trajectory of the instanton solution.
}
\end{figure}
\begin{figure}[t]
\includegraphics[scale=0.7]{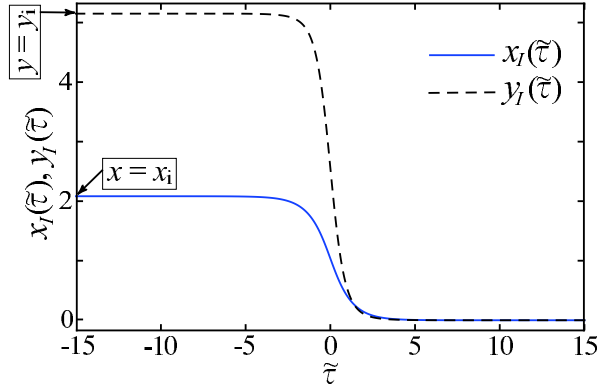}
\caption{\label{fig:ins2D}
Instanton solution $x_{\rm I}(\tilde{\tau})$ and
$y_{\rm I}(\tilde{\tau})$ for $\theta=\pi/L$ and $L=8$.
}
\end{figure}

After considering a toy single-variable approximation to the instanton solution, in this section, we make the next step by increasing the total number of the collective variables to two. Specifically to describe the instanton action we take two variables $\alpha$ and $\beta$ describing the phase slip as independent and for the rest use the linear interpolating function. The phases of such instanton solution (again centered between $1{\rm st}$ and $L$-th sites) are described as
\begin{eqnarray}
\phi_j(\tilde{\tau}) =
\left\{\begin{array}{lll}
\alpha(\tilde{\tau})/2, & {\rm for} & j=1 \\
\alpha(\tilde{\tau})/2+\beta(\tau)+\varphi(\tilde{\tau})(j-2), & {\rm for} & 2\leq j \leq L-1 \\
-\alpha(\tilde{\tau})/2, & {\rm for} & j=L
\end{array}\right. ,
\label{eq:intpl2}
\end{eqnarray}
where $\alpha$ denotes the phase difference between the 1st and $L$-th sites, $\beta$ is
the phase difference between the 2-nd and 1-st (as well as $L$-th and $(L-1)$-th) sites, and
$\varphi \equiv -(\alpha+2\beta)/(L-3)$ is chosen such that the boundary condition $\phi_{L-1} = -\phi_2$ is fulfilled.
Substituting Eq.~(\ref{eq:intpl2}) into Eq.~(\ref{eq:sDless}) we find that the action is described by the two variables $\alpha$ and $\beta$ as
\begin{eqnarray}
\tilde{s}[\alpha,\beta] = \int d\tilde{\tau} \left[
\frac{1}{2} C_{11}\left(\frac{\partial \alpha}{\partial \tilde{\tau}}\right)^2
+ C_{12}\frac{\partial \alpha}{\partial \tilde{\tau}}\frac{\partial \beta}{\partial \tilde{\tau}}
+ \frac{1}{2} C_{22}\left(\frac{\partial \beta}{\partial \tilde{\tau}}\right)^2
+ V_{2}(\alpha, \beta) \right],
\end{eqnarray}
where
\begin{eqnarray}
C_{11}=\frac{L^2+3L+16}{12(L-3)}, \,\,\,
C_{22}=2C_{12}=\frac{(L-1)(L-2)}{3(L-3)},
\end{eqnarray}
and $V_2(\alpha,\beta)$ is the following effective potential
\begin{eqnarray}
V_2(\alpha,\beta) = -2\cos\left(\alpha - \theta \right) -4\cos\left(\beta - \theta \right)
                                   -2(L-3)\cos(\varphi - \theta).
\end{eqnarray}
It is convenient to perform a linear transformation $(x,y)^{\bf t}=\hat{X}(\alpha,\beta)^{\bf t}$, where $\hat{X}$ is an orthogonal $2\times 2$ matrix, to diagonalize the kinetic energy part of the action leading to
\begin{eqnarray}
\tilde{s}[x,y] = \int d\tilde{\tau} \left[
\frac{1}{2} M_x\left(\frac{\partial x}{\partial \tilde{\tau}}\right)^2
+ \frac{1}{2} M_y\left(\frac{\partial y}{\partial \tilde{\tau}}\right)^2
+ V_{2}(x, y) \right].
\label{eq:s2D}
\end{eqnarray}
%

The shape of $V_2(x,y)-V_2(0,0)$ for $\theta=\pi/L$ and $L=8$ is depicted in Fig.~\ref{fig:2Dpote}.
In the potential there are two minima corresponding to the current states with $n=0$ and $n=1$.
The classical equations of motion corresponding to this action are
\begin{eqnarray}
-M_x\frac{\partial^2 x}{\partial\tau^2}+\frac{\partial V_2}{\partial x} =0, \nonumber\\
-M_y\frac{\partial^2 y}{\partial\tau^2}+\frac{\partial V_2}{\partial y} =0.
\label{eq:2Deqm}
\end{eqnarray}
As in the case of the single collective variable, the instanton solution describes the classical trajectory in the inverted potential, which starts from one of the maxima of $-V_2(x,y)$ (or equivalently minima of $V_2(x,y)$) at $-\tilde{\tau}=\tilde{\beta}/2$ and reaches the other maximum at $\tilde{\tau}=\tilde{\beta}/2$ through a saddle point as shown in Fig.~\ref{fig:ins2D}. The trajectory of the instanton solution is indicated by the solid line in Fig.~\ref{fig:2Dpote}. Inserting the instanton solution into Eq.~(\ref{eq:s2D}), we obtain $\tilde{s}_I$.
The derivation of the coefficient $K$ in Eq.~(\ref{eq:espIns}) is almost the same as
that for the single collective variable and we skip it to avoid redundancy. The values of both $\tilde{s}_I$ and $K$ in this two-variable case can be found in the second line of Table~\ref{tab:sK}. In a similar way one can keep on the number of independent degrease of freedom in the instanton solution. 

\subsection{C. All degrees of freedom}
\label{sec:all}
\begin{figure}[b]
\includegraphics[scale=0.7]{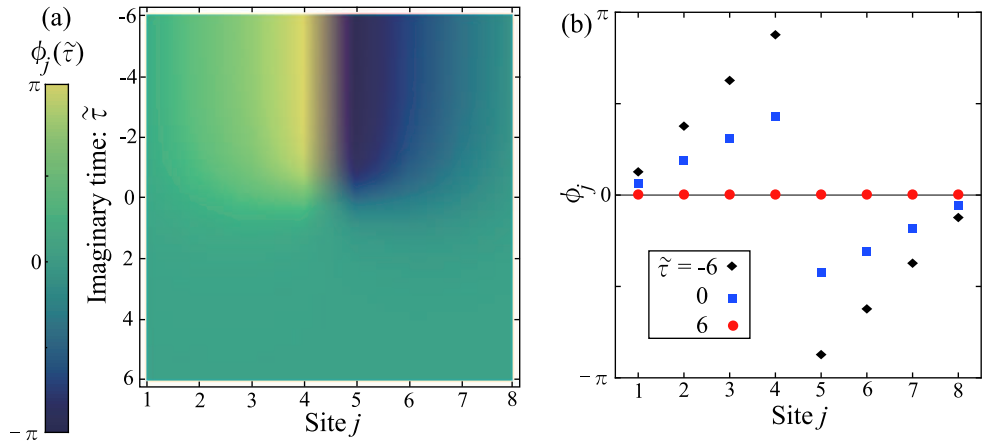}
\caption{\label{fig:ins8D}
(a) Instanton solution $\vec{\phi}_I(\tilde{\tau})$ for $\theta=\pi/L$ and $L=8$.
(b) Snap shots of $\vec{\phi}_I(\tilde{\tau})$ for $\tilde{\tau}=-6$ (black diamonds), $0$
(blue squares), and $6$ (red circles).
}
\end{figure}
As a final step we will explicitly show generalization of the instanton method to the action of the quantum rotor
model Eq.~(\ref{eq:sDless}) where all phases are treated as independent variable. We will show that the energy splitting is given by Eq.~(\ref{eq:A3}) where the eigenvalue equations (\ref{eq:eigEq1}) and (\ref{eq:eigEq2}) are appropriately generalized. For convenience, we rewrite Eq.~(\ref{eq:sDless}) as
\begin{eqnarray}
\tilde{s} = \int d\tilde{\tau} \left[
\frac{1}{2}\frac{\partial \vec{\phi}}{\partial \tilde{\tau}}\cdot \frac{\partial \vec{\phi}}{\partial \tilde{\tau}}
+V(\vec{\phi})
\right]
\label{eq:sVec}
\end{eqnarray}
where $\vec{\phi}$ is an $L$-dimensional vector defined as
\begin{eqnarray}
\vec{\phi} =
\left(
\phi_{1}(\tilde{\tau}),\ldots, \phi_{j}(\tilde{\tau}), \ldots, \phi_{L}(\tilde{\tau})
\right)^{\bf t}
\end{eqnarray}
and the potential is
\begin{eqnarray}
V(\vec{\phi}) = \sum_{j=1}^{L} V_j(\phi_{j+1},\phi_j)
= \sum_{j=1}^{L}  -2\cos\left( \phi_{j+1} - \phi_j - \theta \right).
\end{eqnarray}
The classical equations of motion Eq.~(\ref{eq:classical}) have an instanton solution
$\vec{\phi}(\tilde{\tau}) = \vec{\phi}_I(\tilde{\tau})$ that connects two current states
through the saddle point having a phase kink. We  obtain such a solution by numerically solving Eq.~(\ref{eq:classical}) imposing the boundary conditions:
\begin{eqnarray}
\phi_j(-\tilde{\beta}/2) = \frac{2\pi j}{L}-\pi\left(1+\frac{1}{L}\right),
\,\,\,
\phi_j(\tilde{\beta}/2) = 0.
\label{eq:insBound}
\end{eqnarray}
The corresponding instanton solution for $L=8$ is depicted in Fig.~\ref{fig:ins8D}. Notice that apart from the kink between $4$-th and $5$-th sites the remaining phases approximately linearly depend on the site index justifying the single-variable variational ansatz made in Sec.~\ref{sec:one}. However, because of high sensitivity of the splitting $\Delta$ to especially the value of $\tilde s$ such ansatz can not be used for accurate quantitative calculations. We intentionally shifted the position of the kink in Fig.~\ref{fig:ins8D} to the middle of the system for better graphical presentation. For computational purposes it is convenient to assume that the link develops between 1st and $L$-th sites as we did in earlier calculations. Substituting $\vec{\phi}(\tilde{\tau}) = \vec{\phi}_I(\tilde{\tau})$ into Eq.~(\ref{eq:sVec}), we obtain the instanton action $\tilde{s}_I$.

To calculate $A$ of Eq.~(\ref{eq:capia}), we substitute
\begin{eqnarray}
\vec{\phi}(\tilde{\tau})=\vec{\phi}_I(\tilde{\tau}) + \sqrt{h_{\rm e}}\sum_m c_m \vec{\xi}_m(\tilde{\tau}),
\end{eqnarray}
into $Z_1$ and
\begin{eqnarray}
\vec{\phi}(\tilde{\tau})=\vec{\phi}(-\tilde{\beta}/2)
                                       + \sqrt{h_{\rm e}}\sum_m c_m \vec{\xi}_m^{(0)}(\tilde{\tau}),
\end{eqnarray}
into $Z_0$, where
\begin{eqnarray}
\vec{\xi}_m =
\left(
\xi_{1,m}(\tilde{\tau}),\ldots, \xi_{j,m}(\tilde{\tau}), \ldots, \xi_{L,m}(\tilde{\tau})
\right)^{\bf t},
\end{eqnarray}
\begin{eqnarray}
\vec{\xi}_m^{(0)} =
\left(
\xi_{1,m}^{(0)}(\tilde{\tau}),\ldots, \xi_{j,m}^{(0)}(\tilde{\tau}), \ldots, \xi_{L,m}^{(0)}(\tilde{\tau})
\right)^{\bf t}.
\end{eqnarray}
The $L$ dimensional vectors $\vec{\xi}_m$'s and $\vec{\xi}_m^{(0)}$'s obey
the eigenvalue equations
\begin{eqnarray}
\hat{{\cal M}}\vec{\xi}_m(\tilde{\tau})=\lambda_m\vec{\xi}_m(\tilde{\tau}),
\label{eq:eigEqM1}
\end{eqnarray}
\begin{eqnarray}
\hat{{\cal M}}^{(0)}\vec{\xi}_m^{(0)}(\tilde{\tau})=\lambda_m^{(0)}\vec{\xi}_m^{(0)}(\tilde{\tau}),
\label{eq:eigEqM2}
\end{eqnarray}
and the orthonormalization conditions
\begin{eqnarray}
\int d\tilde{\tau} \,\, \vec{\xi}_l \cdot \vec{\xi}_m = \delta_{l,m},
\,\,\,
\int d\tilde{\tau} \,\, \vec{\xi}_l^{(0)} \cdot \vec{\xi}_m^{(0)} = \delta_{l,m}.
\end{eqnarray}
The $L\times L$ dimensional matrices $\hat{{\cal M}}$ and $\hat{{\cal M}}^{(0)}$ are determined by the matrix elements
\begin{eqnarray}
{\cal M}_{j,k}=\delta_{j,k}
\left(
-\frac{\partial^2}{\partial \tau^2}
+ \left. \frac{\partial^2 V_j}{\partial \phi_j^2} \right|_{\vec{\phi}=\vec{\phi}^I}
+ \left. \frac{\partial^2 V_{j-1}}{\partial \phi_j^2} \right|_{\vec{\phi}=\vec{\phi}^I}
\right)
+ \delta_{j, k-1}
\left. \frac{\partial^2 V_j}{\partial \phi_j \partial \phi_{j+1}} \right|_{\vec{\phi}=\vec{\phi}^I}
+ \delta_{j, k+1}
\left. \frac{\partial^2 V_{j-1}}{\partial \phi_j \partial \phi_{j-1}} \right|_{\vec{\phi}=\vec{\phi}^I},
\end{eqnarray}
\begin{eqnarray}
{\cal M}_{j,k}^{(0)}=\delta_{j,k}
\left(
-\frac{\partial^2}{\partial \tau^2}
+2 \omega^2
\right)
- \delta_{j, k-1} \omega^2
- \delta_{j, k+1} \omega^2,
\end{eqnarray}
where $\omega^2= \left. \partial_{\phi_j}^2 V_j \right|_{\vec{\phi}=\vec{0}}$. Notice that $(L+1)$-th and $0$th sites are equivalent to 1st and $L$-th sites, respectively, reflecting the periodicity of the system. Neglecting the terms higher than the second order with respect to $\sqrt{h_{\rm e}}$, $A$ is again approximated as Eq.~(\ref{eq:A2}).
The derivation of Eq.~(\ref{eq:espIns}) with the coefficient $K$ given by Eq.~(\ref{eq:K}) is exactly the same as the case of the single collective variable and will not be repeated here. The only difference with the single variable case is that $\lambda_m$'s and $\lambda_m^{(0)}$'s are now given by the eigenvalues of Eqs.~(\ref{eq:eigEqM1}) and (\ref{eq:eigEqM2}).

\subsection{D. Comparison with the TEBD results}
\begin{table}[tb]
\begin{center}
\begin{tabular}{| c | c | c |}
\hline
Number of collective & Instanton & Coefficient:  \\
valiables: $m$ & action: $\tilde{s}_I$ & $K$ \\
\hline
1 & \,\,\, 7.749 \,\,\, & \,\,\, 3.71 \,\,\, \\
2 & \,\,\, 7.396 \,\,\, & \,\,\, 4.89 \,\,\, \\
3 & \,\,\, 7.364 \,\,\, & \,\,\, 4.41 \,\,\, \\
4 & \,\,\, 7.363 \,\,\, & \,\,\, 3.64 \,\,\, \\
8 & \,\,\, 7.363 \,\,\,  & \,\,\, 3.06 \,\,\, \\
\hline
\end{tabular}
\end{center}
\caption{\label{tab:sK}
$\tilde{s}_I$ and $K$ for several values of the number of collective variables,
where $L=8$.
}
\end{table}
\begin{figure}[b]
\includegraphics[scale=0.6]{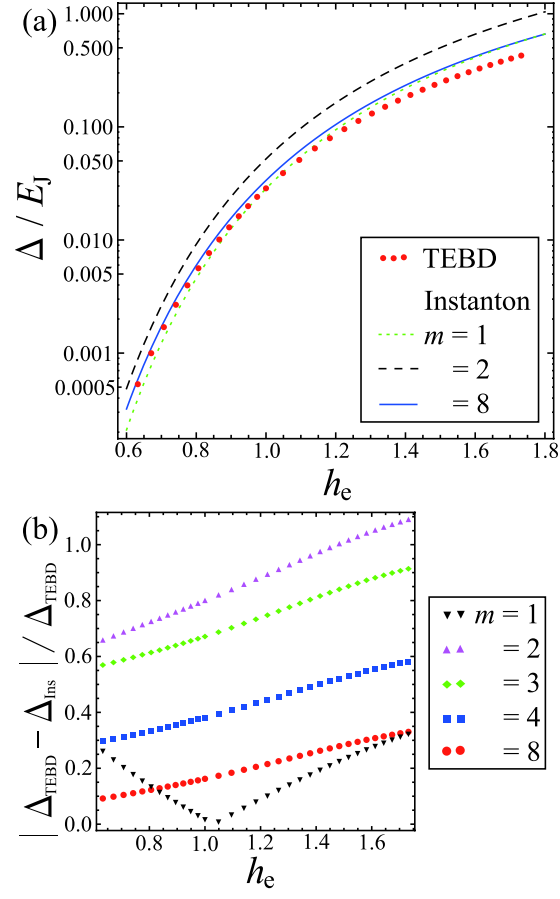}
\caption{\label{fig:EspC}
(a) Energy splitting $\Delta/E_{\rm J}$ versus the effective Planck's constant
$h_{\rm e}\equiv \sqrt{U/(\nu J)}$.
The dotted, dashed, and solid lines represent the results by the instanton method for
$m = 1, 2$, and $8$.
The dots are the TEBD results for $\nu = 1000$.
(b) Ratio $|\Delta_{\rm TEBD} - \Delta_{\rm Ins}|/\Delta_{\rm Ins}$ of the difference
between the TEBD and instanton results as a function of $h_{\rm e}$.
}
\end{figure}
In the main text we compared the energy splitting calculated by the instanton method
with all possible degrees of freedom (as described in Sec.~\ref{sec:all}) with the TEBD results. It is also instructive to learn how the instanton method is improved as the number of collective variables increases. For this purpose, let us now present the comparison between the TEBD and the approximate instanton results where only $m<L$ collective variables are treated as independent. We take a relatively small system size $L=8$. In Table~\ref{tab:sK}, we first show the instanton action $\tilde{s}_I$ and the coefficient $K$ for several values of $m$. Both $\tilde{s}_I$ and $K$ approache the exact values corresponding to $m = 8$ as $m$ increases. We note that the action $\tilde{s}_I$ for $m = 4$ is exactly the same as that for $m = 8$ because the instanton solution is anti-symmetric with respect to $j\to L-j$, i.e. with respect to the link at which the phase kink develops (see Fig.~\ref{fig:ins8D}). In contrast, $K$ for $m = 4$, where the fluctuations are also forced to obey the same symmetry as well, is significantly different from $K$ for $m = 8$.
Thus, it is crucial to include all possible fluctuations in order to obtain the correct value of $K$.

In Fig.~\ref{fig:EspC}(a), we plot the energy splitting calculated by the instanton method
as a function of $h_{\rm e}$ together with that by TEBD for $\nu = 1000$. At first glance, it seems that the results by the instanton method with a single collective variable agrees very well with the TEBD results. However, this seeming agreement is rather coincidental as shown in Fig.~\ref{fig:EspC}, where we plot the ratio $|\Delta_{\rm TEBD} - \Delta_{\rm Ins}|/\Delta_{\rm Ins}$ of the difference between the energy splittings by TEBD, $\Delta_{\rm TEBD}$, and the instanton method, $\Delta_{\rm Ins}$. There we clearly see that the error for $m = 1$ (black triangles) does not monotonically decrease with $h_{\rm e}$, contradicting the basic fact that the instantons should be more accurate at smaller $h_{\rm e}$. Except for this case with $m= 1$, the error decreases monotonically as the number of independent phases $m$ increases and the effective Planck's constant $h_{\rm e}$ decreases.

\end{document}